\documentclass[twocolumn,aps,showpacs,nofootinbib,superscriptaddress,prl,preprintnumbers]{revtex4}
\usepackage{mathrsfs}
\usepackage{epsfig}
\usepackage{textcomp}
\usepackage{amsmath}
\usepackage{graphicx}
\usepackage{amssymb}
\usepackage{color}
\usepackage{appendix}
\usepackage{url}

\newcommand{\llep}{\ell^{+}\ell^{-}}
\begin{document}

\title{Probing the Higgs boson-gluon coupling via the jet energy profile at $e^+e^-$ colliders}

\author{Gexing Li}
\email{ligx@ihep.ac.cn}
\affiliation{Institute of High Energy Physics, Chinese Academy of Sciences, Beijing 100049, China}
\affiliation{School of Physics Sciences, University of Chinese Academy of Sciences, Beijing 100039, China}
\author{Zhao Li}
\email{zhaoli@ihep.ac.cn}
\affiliation{Institute of High Energy Physics, Chinese Academy of Sciences, Beijing 100049, China}
\affiliation{School of Physics Sciences, University of Chinese Academy of Sciences, Beijing 100039, China}

\author{Yandong Liu}
\email{ydliu@bnu.edu.cn}
\affiliation{Key Laboratory of Beam Technology of Ministry of Education, College of Nuclear Science and Technology, Beijing Normal University, Beijing 100875, China}
\affiliation{Beijing Radiation Center, Beijing 100875, China}

\author{Yan Wang}
\email{wangyan728@ihep.ac.cn}
\affiliation{Institute of High Energy Physics, Chinese Academy of Sciences, Beijing 100049, China}
\affiliation{Deutsches Elektronen-Synchrotron, Hamburg 22607, Germany}

\author{Xiaoran Zhao}
\email{xiaoran.zhao@uclouvain.be}
\affiliation{Centre for Cosmology, Particle Physics and Phenomenology (CP3), Universit\'{e} catholique de Louvain, 1348 Louvain-la-Neuve, Belgium}


\preprint{MCnet-18-09, CP3-18-28}

\begin{abstract}
The effective coupling of the Higgs boson to a gluon pair is
one of the most important parameters to test the Standard Model and search for the new physics beyond.
In this paper, we propose several new observables based on the jet energy profile to extract the effective coupling.
The statistical uncertainties of the effective coupling extracted by using new observables are derived and estimated based
on the simulation at the future $e^+e^-$ collider for $250$ GeV center-of-mass energy and 5 ab$^{-1}$ integrated luminosity.
We found that the statistical uncertainties of effective coupling via the optimized observable can reach about $1.6\%$
in the channels of a $Z$ boson decaying to lepton pairs and is reduced by $52\%$ compared to the relevant uncertainties in the conventional approach. These new observables potentially can be helpful for the measurement of effective coupling at future $e^+e^-$ colliders.

\end{abstract}

\maketitle

\section{Introduction}

The discovery of the Higgs boson at the CERN Large Hadron Collider (LHC) has marked the completeness and success of the Standard Model (SM).
Thereafter, the precision measurement of the properties of Higgs boson has become the most promising approach to completely
understanding the Higgs mechanism, since the SM predicts not only one scalar boson but also the couplings of the Higgs boson to
the other SM particles.
The effective coupling of the Higgs boson to a gluon pair is one of the most important
parameters to test the Standard Model and thus to search for the new physics beyond,
since it can be directly affected by the new physics particle
loop \cite{Einhorn:1993hj,Kanemura:2004mg,He:2013tia,Moyotl:2016fdk,Baek:2017kxh,Hou:2017vvp,Kanemura:2017wtm,Passehr:2017ufr}.

The measurement of Higgs boson-gluon effective coupling is mainly via gluon fusion at the LHC \cite{Khachatryan:2016vau,CMS:2018lkl}.
However, the overwhelmingly large QCD background hinders the precise search for this process.
And different Higgs couplings are mixed together in a process of Higgs production and decay, which leads to the Higgs boson-gluon effective coupling being affected by the uncertainties of other Higgs couplings.
But the environment of an electron-positron collider is very clean, and the main process of Higgs production is the Higgsstrahlung process $e^+e^- \rightarrow Zh$. The measurement of cross section $\sigma_{Zh}$ is independent of the Higgs decay mode by the $Z$ boson recoil mass method, which allows us to solely extract the Higgs boson-gluon effective coupling from Higgs decay.
Therefore, the next generation of electron-positron colliders becomes an inevitable choice \cite{Peskin:2012we,Peskin:2013xra}.

In the past few years, several options have been proposed as a Higgs factory,
for instance, Circular Electron-Positron Collider (CEPC) \cite{CEPC-SPPCStudyGroup:2015csa,CEPC-SPPCStudyGroup:2015esa,Mo:2015mza},
Future Circular Collider-electron-positron \cite{Gomez-Ceballos:2013zzn,Barletta:2014vea, Benedikt:2016vzy},
and International Linear Collider (ILC) \cite{Behnke:2013xla,Adolphsen:2013kya,Behnke:2013lya}.
At the Higgs factory, the measurement of most of the Higgs properties can be expected to reach a high accuracy.
And ideally the Higgs boson-gluon coupling can be investigated by extracting the $gg$ mode in Higgs boson decays.
With $b$-tagging efficiency 80\%, the accuracy of Higgs boson-gluon coupling will reach $2.2\%$
for the channels of a $Z$ boson decaying to a lepton pair before using a template fit,
and can be further improved to $1.5\%$ after using a template fit\cite{YuBai}.

However, according to the SM predictions on the decays of the $125$~GeV Higgs boson \cite{Aad:2015gba, Khachatryan:2016vau},
the $gg$ mode has a very small branching ratio ${\cal B}^{\rm SM}_{gg}\equiv{\cal B}^{\rm SM}(h\to gg) =8.56\%$
and phenomenologically manifests dijet signals.
Meanwhile, the $b\bar{b}$ mode [${\cal B}^{\rm SM}_{b\bar{b}}\equiv{\cal B}^{\rm SM}(h\to b\bar b)=58.09\%$]
and $c\bar{c}$ mode [${\cal B}^{\rm SM}_{c\bar{c}}\equiv{\cal B}^{\rm SM}(h\to c\bar c)=2.9\%$]
have sizable contributions to the dijet events.
This would be a serious drawback for the efficiency to extract the Higgs boson-gluon coupling.

In view of the experimental observation,
the dijet decay mode of the Higgs boson has a dominant contribution from the bottom quark pair,
so the Higgs boson-gluon coupling is overwhelmed.
To reveal Higgs boson-gluon coupling from Higgs decay, $b$ tagging is an efficient tool to suppress the bottom quark contribution.
However, the $b$ tagging is not enough to fully eliminate the quark jets from Higgs boson decay and the background processes to Higgs production.
Therefore, it is worthy to elaborate other approaches to promote the branching ratio measurement.

One long-standing and extensively studied goal at the collider is how to efficiently distinguish the jets induced by the quark and gluon.
Of all the proposed variables to achieve the goal, the jet energy profile (JEP) is a conventional one.
For a jet of cone size $R$, the JEP is defined as
\begin{align}
\psi (r) = \frac{1}{N_j} \sum_{j} \psi_{j}(r) =
\frac{1}{N_j} \sum_j \frac{\displaystyle{\sum_{r_i < r} p_{{\rm T},i}(r_i)   }}{\displaystyle{\sum_{r_i<R}  p_{{\rm T},i}(r_i)  }},
\label{averageJEP}
\end{align}
where $r$ ($\le R$) is the size of a test cone.
$N_j$ is the total number of jets.
$p_{{\rm T},i}$ and $r_i$ are the transverse momentum and the distance from the jet axis of the $i$th constituent, respectively.
And $\psi_j(r)$ represents the JEP of a single jet, so $\psi(r)$ can also be defined as the average JEP of jets.
Generally, a gluon jet has a different JEP shape from a quark jet due to more QCD radiation.
Since the usually observed jets are the mixing of quark jets and gluon jets,
the overall JEP would be the weighted average of the quark-jet JEP and gluon-jet JEP,
and its shape can imply the ratio between quark jets and gluon jets.
Many works have utilized the JEP to improve the analysis, for instance,
identifying Higgs production mechanisms \cite{Rentala:2013uaa}, searching for dark matter interactions \cite{Agrawal:2013hya},
and detecting new physics in dijet resonance \cite{Chivukula:2015tca}.

In this paper, we assume the new physics influences only the Higgs boson-gluon coupling
and can be summarized into the effective operator of a Higgs boson-gluon-gluon interaction\cite{He:2013tia}:
\begin{equation}
{\cal L}_{hgg} = \kappa_g c_{\rm SM}^g \frac{\alpha_s}{12\pi v} hG_{\mu\nu}^a G^{a\mu\nu},
\label{effectiveOperator}
\end{equation}
where $c_{\rm SM}^g$ is the SM prediction of Higgs boson-gluon effective coupling from a heavy quark loop.
$\kappa_g$ represents the deviation from the SM prediction, i.e., $\kappa_g = 1$ in the SM.
By analyzing the dijet decay mode of the Higgs boson that is produced via the process $e^+ e^- \rightarrow Z h$ at the future $e^+e^-$ collider,
instead of the conventional averaged JEP shown in Eq. (\ref{averageJEP}), we extract the information of $\kappa_g$ from the accumulated JEP,
which has better sensitivity to $\kappa_g$ and will be explained in detail in the next section.
In the analysis the $b$ tagging and $c$ tagging are included to suppress the contribution from bottom pair and charm pair decay modes.

The content is organized as follows. In the next section, several observables are defined based on the JEP,
and the relevant uncertainties of $\kappa_g$ via different observables are derived.
In the third section, the Monte Carlo (MC) simulation including background events
is used for a comparison between different observables at a future $e^+e^-$ collider.
Then a conclusion is made in the final section.

\section{Accumulated Jet Energy Profile}

Suppose the new physics beyond the SM could modify the Higgs boson-gluon effective operator as shown in Eq. (\ref{effectiveOperator});
it affects the decay branching ratio ${\cal B}_{gg} = \kappa_g^2 {\cal B}^{\rm SM}_{gg}$ \cite{Khachatryan:2016vau,CMS:2018lkl,CEPC-SPPCStudyGroup:2015csa,CEPC-SPPCStudyGroup:2015esa}.
Therefore, by the definition in Eq. (\ref{averageJEP}), the energy profile of jets from the Higgs dijet decay channel can be explicitly expressed as
\begin{align}
\psi(r)=
\frac{ \kappa_g^{2}{\cal B}^{\rm SM}_{gg} \psi_g
+{\cal B}^{\rm SM}_{q\bar{q}}\psi_q
}
{ \kappa_g^{2}{\cal B}^{\rm SM}_{gg}
+{\cal B}^{\rm SM}_{q\bar{q}}
},
\label{JEPtag}
\end{align}
where $\psi_g$ and $\psi_q$ are the energy profiles of gluon jet and quark jet, respectively. And
\begin{align}
{\cal B}^{\rm SM}_{q\bar{q}} \equiv {\cal B}^{\rm SM}_{b\bar{b}}(1-\varepsilon_b)^{2}+{\cal B}^{\rm SM}_{c\bar{c}}(1-\varepsilon_c)^{2}.
\end{align}
Here, the quark jet is composed of a charm jet and a bottom jet,
and the JEP of the quark-jet is obtained by the weighted average of the JEP of the charm jet and bottom jet.
Meanwhile, in order to increase the sensitivity of the JEP to $\kappa_g$,
both $b$ tagging and $c$ tagging ($b$ and $c$ tagging) have been applied to suppress the contributions from the bottom jet and charm jet.
And $\varepsilon_b$ and $\varepsilon_c$ are the efficiencies of $b$ tagging and $c$ tagging, respectively.

In the above equation, the decay branching ratios ${\cal B}^{\rm SM}_{gg}$, ${\cal B}^{\rm SM}_{b\bar{b}}$, and ${\cal B}^{\rm SM}_{c\bar{c}}$
can be obtained by the SM predictions, and the JEP of the quark jet and gluon jet, $\psi_q$ and $\psi_g$, can be obtained
by a MC simulation or perturbative QCD prediction\cite{Li:2011hy}.

\begin{figure}[!htb]
\includegraphics[scale=0.37]{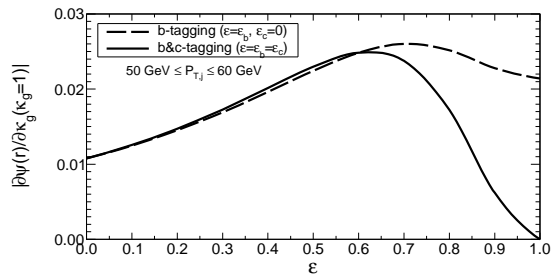}
\caption{\label{FIG:1}
The slopes of the JEP with respect to $\kappa_g$ at $\kappa_g=1$ as a function of tagging efficiency $\varepsilon$
after $b$ tagging($\varepsilon=\varepsilon_{b}, \varepsilon_{c}=0$) and $b$ and $c$ tagging($\varepsilon=\varepsilon_{b}=\varepsilon_{c}$).
The transverse momentum of jets 50 GeV $\leq P_{T,j} \leq$ 60 GeV.
The test cone size $r=0.3$.
The jets from the background to $e^+e^-\to Zh\to Zjj$ are not included yet in order to demonstrate the physics more clearly.
}
\end{figure}
\begin{figure}[!htb]
\includegraphics[scale=0.37]{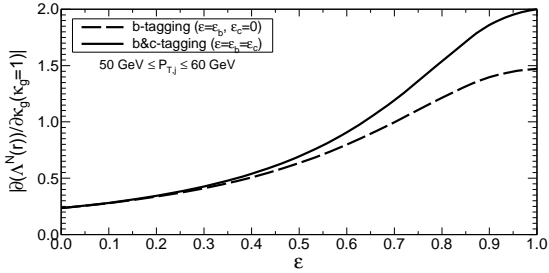}
\caption{\label{FIG:2}
The slope of $\Lambda^{N}(r)$ at $\kappa_g=1$ to the efficiency after $b$ tagging($\varepsilon=\varepsilon_{b}, \varepsilon_{c}=0$) and $b$ and $c$ tagging($\varepsilon=\varepsilon_{b}=\varepsilon_{c}$).
The transverse momentum of jets 50 GeV $\leq P_{T,j} \leq$ 60 GeV. The test cone size $r=0.3$. }
\end{figure}

It can be found that in Eq. (\ref{JEPtag}) a conservative choice of tagging efficiency ($\varepsilon = \varepsilon_b=\varepsilon_c\approx 70\%$)
can decrease the contribution of the bottom jet and charm jet to the same size as that of the gluon jet,
so the extraction of $\kappa_g$ could become more efficient.
However, as shown in Fig.\ref{FIG:1}, while the tagging efficiency becomes better than $70\%$,
the sensitivity of the JEP to the Higgs boson-gluon coupling $\kappa_g$ will start getting worse.
And when the tagging is ideally perfect ($\varepsilon=100\%$), the JEP becomes independent of $\kappa_g$,
which is inconvenient for the extraction of $\kappa_g$ from the JEP.
This behavior can be understood in Eq. (\ref{JEPtag}), where the $b$ and $c$ tagging would suppress the contribution from the bottom jet and charm jet
so that the $\kappa_g$ in the denominator is revealed to just cancel with the one in the numerator.
One of the direct solutions is to use an accumulated JEP, which does not contain $\kappa_g$ in the denominator.
Therefore, we define the new observable based on the accumulated JEP as
\begin{equation}
\Lambda^N (r) \equiv \frac{ \sum_j \psi_j(r) }{ \sum_j^{\rm SM} \psi_j(r) }.
\end{equation}
For the dijet decay channel of a Higgs boson including $b$ and $c$ tagging,
it can be explicitly expressed as
\begin{align}
\Lambda^N (r)=
\frac{
\kappa_g^{2}{\cal B}^{\rm SM}_{gg} \psi_g
+{\cal B}^{\rm SM}_{q\bar{q}}\psi_q
}
{
{\cal B}^{\rm SM}_{gg} \psi_g
+{\cal B}^{\rm SM}_{q\bar{q}}\psi_q
}.
\label{LambdaHiggsJet}
\end{align}
Now it can be seen that in the ideal condition the perfect tagging
can directly simplify this observable $\Lambda^N(r) = \kappa_g^2$.
In Fig. \ref{FIG:2}, we plot the slope of $\Lambda^N(r)$ as the sensitivity to $\kappa_g$
as a function of the tagging efficiency for the test cone size $r=0.3$.
This figure shows that the sensitivity of observable $\Lambda^N(r)$ to $\kappa_g$
keeps increasing as the tagging becomes better in the whole region, as expected.

Although the above analysis has shown a promising measurement on $\kappa_g$ after including $b$ and $c$ tagging,
the background contamination has not been included yet and the light quark jet cannot be easily vetoed by $b$ and $c$ tagging.
Therefore, after including the background to Higgs production, Eq. (\ref{LambdaHiggsJet}) can be extended into
\begin{align}
\Lambda^{N}(r)=
\frac{
\kappa_g^{2}\sigma_{h}{\cal B}^{\rm SM}_{gg} \psi_g
+\sigma_{h}{\cal B}^{\rm SM}_{q\bar{q}}\psi_q
+\sigma_{jj}^{\rm BG} \psi_{j}^{\rm BG}
}{
\sigma_{h}{\cal B}^{\rm SM}_{gg} \psi_g
+\sigma_{h}{\cal B}^{\rm SM}_{q\bar{q}}\psi_q
+\sigma_{jj}^{\rm BG} \psi_{j}^{\rm BG}
},
\label{LambdaAllJet}
\end{align}
where $\sigma_h $ is the Higgs production rate at the Higgs factory.
$\sigma_{jj}^{\rm BG}$ is the production rate of background events \cite{XinMo}.
$\psi_{j}^{\rm BG}$ includes the jets from the background.
Although the jets from the background explicitly reduce the sensitivity to $\kappa_g$ in Eq.(\ref{LambdaAllJet}),
we found the JEP in each term plays the role of weight for each contribution to $\Lambda^N$.
Therefore, these JEP weights can be shifted to increase the sensitivity to $\kappa_g$.
For instance, if the JEP is subtracted simultaneously by the average JEP of quark jets from Higgs decay and jets from the background,
i.e., $\tilde\psi = (\sigma_{h}{\cal B}^{\rm SM}_{q\bar{q}}\psi_q+\sigma_{jj}^{\rm BG} \psi_{j}^{\rm BG})/(\sigma_{h}{\cal B}^{\rm SM}_{q\bar{q}}+\sigma_{jj}^{\rm BG})$,
ideally one can obtain the most sensitive measurement on $\kappa_g$.
However, in practice, this subtraction could not be perfect, and the uncertainty of $\kappa_g$ may not be optimized.
Therefore, by simultaneously shifting the JEP, we define a generic observable
\begin{align}
    Z^N(r) = \frac{ \sum_j (\psi_j+a) }{ \sum_j^{\rm SM} (\psi_j+a) },
\end{align}
where $a$ is a tunable parameter.

After including the background contribution, it will be necessary to understand the uncertainty of $\kappa_g$ extracted from the new observables.
As an intermediate quantity, the uncertainties of the new observables include the statistical and systematic uncertainties.
The evaluation of systematic uncertainties requires a detailed detector study and is unknown yet for the Higgs factory.
However, the statistical uncertainties can be obtained from a MC simulation.
Explicitly, the new observable $Z^{N}(r)$ can be written as
\begin{align}
Z^{N}(r)=[N_{g}(\psi_g+a)+N_{q}(\psi_q+a)+N_{\rm BG}(\psi_{\rm BG}+a)]\big/C^{\rm SM},
\end{align}
where the normalization factor
\begin{align}
C^{\rm SM}
=N_{g}^{\rm SM}(\psi_g+a)+N_{q}^{\rm SM}(\psi_q+a)+N_{\rm BG}^{\rm SM}(\psi_{\rm BG}+a).
\end{align}
$N_{g}$, $N_{q}$, and $N_{\rm BG}$ are, respectively, the number of gluon jets from Higgs boson decay, quark jets from Higgs boson decay, and jets from the background.
The total number of jets $N=N_{g}+N_{q}+N_{\rm BG}$.

Then the uncertainty of $Z^{N}(r)$ is
\begin{align}
\delta Z^{N}(r)=&\Big[N \sigma^{2}(r)+N_{g}(\psi_g+a)^{2}+N_{q}(\psi_q+a)^{2}
\nonumber \\ &
+N_{\rm BG}(\psi_{\rm BG}+a)^{2}\Big]^{1/2}\big/C^{\rm SM}.
\end{align}
The first term is a fluctuation of the JEP,
which is $N\sigma^{2}(r)=N_{g}^{2}(\delta \psi_{g})^{2}+N_{q}^{2}(\delta \psi_{q})^{2}+N_{\rm BG}^{2}(\delta \psi_{\rm BG})^{2}$.
The other three terms are fluctuations of relevant event numbers.

Meanwhile, the uncertainty on the measurement of $Z^N$ will be passed to the uncertainty on $\kappa_g$ via the following formula:
\begin{align}
\delta \kappa_g^Z=
\delta Z^N \left|\frac{\partial Z^N}{\partial \kappa_g}\right|^{-1},
\end{align}
where the superscript $Z$ indicates that this uncertainty is obtained by measuring observable $Z^N$.

Then the uncertainty of $\kappa_g$ around the SM prediction $\kappa_g=1$ can be explicitly expressed as
\begin{align}
    \delta \kappa_g^Z= &
    \delta \kappa_g^N
\Big[\left(\frac{\sigma(r)}{\psi_g+a}\right)^{2}+f_{g}+f_{q}\left(\frac{\psi_q+a}{\psi_g+a}\right)^{2}
\nonumber \\ &
+f_{\rm BG}\left(\frac{\psi_{\rm BG}+a}{\psi_g+a}\right)^{2}\Big]^{1/2}.
\label{errorZ}
\end{align}
where the factor $\delta \kappa_g^N=\sqrt{N}/2 N_g$ is the statistical uncertainty
of $\kappa_g$ via the conventional approach and the $f_{g}$, $f_{q}$, and $f_{\rm BG}$ are,
respectively, the fraction of gluon jets from Higgs boson decay, quark jets from
Higgs boson decay, and jets from the background with respect to the total number of jets.
When $r=R$, the JEP will become unity, $\psi_g = \psi_q = 1$ and $\sigma(r)=0$,
and the observable $Z^{N}$ will be converted to the conventional approach.

By tuning the parameter $a$, we can give a heavier weight to the signal and make the size of the first term controllable. The minimal uncertainty $\delta \kappa_g^Z$ can be met at
\begin{equation}
\frac{\partial \delta \kappa_g^Z}{\partial a} = 0,
\end{equation}
which can provide the solution
\begin{equation}
a= \frac{\sigma^{2}(r) + f_{\rm BG}(\psi_q-\psi_{\rm BG})(\psi_g-\psi_{\rm BG})}
{f_{\rm q}(\psi_g-\psi_q) + f_{\rm BG}(\psi_g-\psi_{\rm BG})}-\psi_q.
\end{equation}
If the background only contributes quark jets, this solution can be simplified as
\begin{equation}
a= \frac{\sigma^{2}(r)}{(\psi_g-\psi_q)f_{\rm B}} - \psi_q,
\end{equation}
where
\begin{align}
f_{\rm B}=(N_b+N_c+N_{\rm BG})/N.
\end{align}
Then
\begin{align}
    \delta \kappa_g^Z= \delta \kappa_g^N \left\{1-f_{\rm B} \left[1+\frac{\sigma^2(r)}{(\psi_g-\psi_q)^2 f_{\rm B}} \right]^{-1} \right\}^{1/2}.
\end{align}

If the background is much bigger than the signal, the fraction $f_{\rm B}$ is approximately equal to 1. The uncertainty of $\kappa_g$ can be simplified as
\begin{align}
    \delta \kappa_g^Z \approx \delta \kappa_g^N {\left[1+\frac{(\psi_g-\psi_q)^2}{\sigma^2(r)}\right]}^{-1/2}.
\end{align}
It shows that the new observable $Z^N(r)$ will get more improvement than the conventional
approach if the difference of the JEP between the quark and gluon is big and the uncertainty of the JEP is small.
Suppose $\sigma(r) \ll \left|\psi_g-\psi_q \right|$; the expansion of $\delta \kappa_g^Z$ is
\begin{align}
    \delta \kappa_g^Z = \delta \kappa_g^N \left[\frac{\sigma(r)}{\left|\psi_g-\psi_q \right|}+\mathcal{O} \right].
\label{errorx}
\end{align}
Equation (\ref{errorx}) that ignores the higher-order terms is equivalent to setting
the parameter $a=-\psi_q$ in Eq.(\ref{errorZ}). Then the $Z^N (a=-\psi_q)$ will
be a perfect observable to reduce the uncertainty of $\kappa_g$ in this case.
However, if $\sigma(r)$ and $\left|\psi_g-\psi_q \right|$ are at the same order
so that the higher-order terms of Eq.(\ref{errorx}) cannot be ignored,
the $Z^N (a=-\psi_q)$ will transfer a quite large uncertainty to $\kappa_g$.

For other specific values of $a$, the observable $Z^N$ will degrade to some simple observables, for example, $\Lambda^{N}=Z^N (a=0)$.
The uncertainty of $\kappa_g$ via measuring observable $\Lambda^{N}$ can be explicitly shown as
\begin{align}
    \delta \kappa_g^\Lambda=&
    \delta \kappa_g^N
\Big[\left(\frac{\sigma(r)}{\psi_g}\right)^{2}+f_{g}+f_{q}\left(\frac{\psi_q}{\psi_g}\right)^{2}
\nonumber \\ &
+f_{\rm BG}\left(\frac{\psi_{\rm BG}}{\psi_g}\right)^{2}\Big]^{1/2}.
\label{errorLambda}
\end{align}
Since the quark jet is usually narrower than the gluon jet, i.e., $\psi_q > \psi_g$,
the ratios $\psi_q / \psi_g$ and $\psi_{\rm BG} / \psi_g$ will give heavier weights
to quark jets from Higgs boson
decay and background jets. Therefore, this observable $\Lambda^{N}$ will be a little worse than the conventional approach.

In order to give heavier weights to gluon jets, another interesting observable is choosing the part of the jet that lies outside the test cone of size $r$, which equals $Y^N=Z^N(a=-1)$:
\begin{equation}
Y^N (r) = \frac{ \sum_j (1-\psi_j) }{ \sum_j^{\rm SM} (1-\psi_j) }.
\end{equation}
Similarly, the uncertainty of $\kappa_g$ via measuring observable $Y^{N}$ can be explicitly shown as
\begin{align}
    \delta \kappa_g^Y=&
    \delta \kappa_g^N
\Big[\left(\frac{\sigma(r)}{1-\psi_g}\right)^{2}+f_{g}+f_{q}\left(\frac{1-\psi_q}{1-\psi_g}\right)^{2}
\nonumber \\ &
+f_{\rm BG}\left(\frac{1-\psi_{\rm BG}}{1-\psi_g}\right)^{2}\Big]^{1/2}.
\label{errorY}
\end{align}
In this observable, the signal will obtain a heavier weight than the background.
Therefore, the observable $Y^N$ is expected to be more sensitive to $\kappa_g$ than observable $\Lambda^N$.

\section{Simulation}

In this section, we will investigate the new observables based on the accumulated JEP
proposed in the previous section by analyzing Standard Model MC events, which are generated by Whizard 1.95 and showered by Pythia 6 at
future $e^+e^-$ colliders \cite{CEPC-SPPCStudyGroup:2015csa,CEPC-SPPCStudyGroup:2015esa,Mo:2015mza,Gomez-Ceballos:2013zzn,Barletta:2014vea, Benedikt:2016vzy,Behnke:2013xla,Adolphsen:2013kya,Behnke:2013lya}
for the center-of-mass energy 250 GeV and integrated luminosity 5 ab$^{-1}$.
The signal events are classified into two channels according
to $Z$ boson decays: $Z\to e^{+}e^{-}$ and $Z\to \mu^{+}\mu^{-}$.
The background events to $Zh$ production are described in Ref. \cite{Mo:2015mza}.

The event reconstruction procedure first selects the isolated leptons with an energy of more than 10 GeV.
The two opposite charged ones of the isolated leptons are selected to reconstruct the $Z$ boson by minimizing\cite{Yan:2016xyx}
\begin{align}
\chi^2(M_{\llep})=\frac{(M_{\llep}-M_{Z})^2}{\sigma^2_{M_{\llep}}}+\frac{(M_\mathrm{h,rec}-M_{h})^2}{\sigma^2_{M_\mathrm{h,rec}}},
\end{align}
where $\sigma_{M_{\llep}}$ and $\sigma_{M_\mathrm{h,rec}}$ are the Gaussian fits to the distribution of $M_{\llep}$ and $M_\mathrm{h,rec}$, respectively.
Here $M_{\mathrm{h,rec}}$ is the recoil mass for the hypothetical Higgs boson via the kinematic relation.
After finding the lepton pair, the rest of the final states are clustered into jets by using anti-$kt$ algorithm \cite{Cacciari:2011ma}
with cone size $R=1.5$, and the energy of every jet is required to be more than 5 GeV.
Clustering jets with a big cone size is helpful to reconstruct Higgs boson at lepton colliders \cite{XinMo}.
Two jets, whose invariant mass is close to the Higgs mass and recoil mass is close to the $Z$ boson mass at the same time, will be selected.
Then the following kinematic cuts are applied to reject the backgrounds:
\begin{itemize}
\item the lepton pair invariant mass $M_{\llep} \in$  [73,120] GeV,
\item the transverse momentum of lepton pair $P_T^{\llep} \in [10,70]$ GeV,
\item the value of gradient boosted decision trees (BDTG) $\in$  [-0.25,1],
\item the lepton pair recoil mass $M_{\mathrm{h,rec}}\in[110,155]$ GeV,
\item the total energy of all the visible particles except the lepton pair $E_\mathrm{vis}>10$ GeV,
\item the polar angle of leading and subleading selected jets $\cos\theta\in [-0.98,0.98]$,
\item the energy of leading selected jet $E_j \geq 45$ GeV,
\item the energy of subleading selected jet $E_{sub.j} \geq 15$ GeV,
\item the two jets invariant mass $M_{jj}\in [95,130]$ GeV, and
\item the two jets recoil mass $M_{jj}\in [85,130]$ GeV.
\end{itemize}
The leptonic cuts include the BDTG cut follow the ILC paper\cite{Yan:2016xyx}. The BDTG input variables are the mass of $Z$ boson $M_{\llep}$, the polar angle of the $Z$ boson $\cos(\theta_{Z})$, the angle between the lepton pair $\cos(\theta_{lep})$, and the polar angle of each lepton track $\cos(\theta_{track,1,2})$. More details about them can be found in Ref. \cite{Yan:2016xyx}. The jet cuts are referred to the work of the CEPC Working Group\cite{YuBai} and tuned to get the optimal results.

The mistag efficiency of the charm quark to the bottom
quark is $\varepsilon_{c \rightarrow b}=10\%$, and the light quark to
the bottom quark is $\varepsilon_{q \rightarrow b}=0$ when the $b$-tagging
efficiency is $\varepsilon_b=80\%$. And the mistag efficiency
of the bottom quark to the charm quark is $\varepsilon_{b \rightarrow c}=12\%$,
and the light quark to the charm quark is $\varepsilon_{q \rightarrow c}=7\%$
when the $c$-tagging efficiency is $\varepsilon_c=60\%$.
Suppose that the gluon has the same mistag efficiency as the light quark \cite{ManqiRUAN}.

The JEP approach contains two aspects: JEP cut and JEP weight. The JEP cut is adding a unique cut which requires the JEP of leading and subleading jets $\psi_j \in [0.05, 0.99]$ in every event. This cut can effectively remove the background jets by analyzing the internal structure of jets and decrease the JEP uncertainty $\sigma(r)$. The JEP weight means quark jets and gluon jets are given different weights by the new observables. And the parameter $a$ can be tuned to give a heavier weight to gluon jets as we have already analyzed in the previous section.

\begin{figure}[!htb]
\includegraphics[scale=0.37]{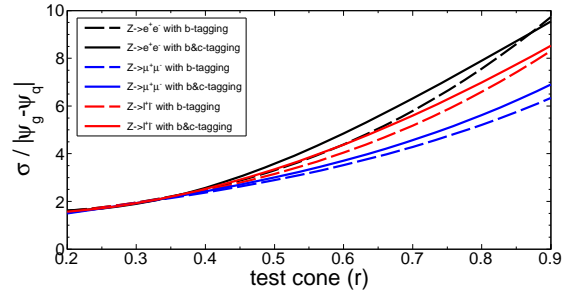}
\caption{\label{FIG:3}
The ratios of JEP uncertainties with respect to the difference of the JEP between quark jets and gluon jets after using the JEP cut varies with the JEP test cone $r$ in different channels of $Z$ boson decay by implementing only $b$ tagging ($\varepsilon_b = 80\%$) (dotted line) and both $b$ tagging ($\varepsilon_b = 80\%$) and $c$ tagging ($\varepsilon_c = 60\%$) (solid line). }
\end{figure}

Figure \ref{FIG:3} shows the ratios of JEP uncertainties $\sigma(r)$ with respect to the difference of the JEP between quark jets and gluon jets $\left|\psi_g-\psi_q \right|$ vary with the JEP test cone $r$.
The ratios in different channels all increase as the test cone increases and have a similar value at test cone 0.2-0.4.
The ratios in the electron channel is higher than that in the muon channel at the test cone region 0.4-0.9, and the gap is growing as the test cone increases.
The ratios in the $Z\rightarrow \ell^{+}\ell^{-}$ channel, which combines the two lepton channels $Z\rightarrow e^{+}e^{-}$ and $Z\rightarrow \mu^{+}\mu^{-}$, are between that in the electron and muon channels.
For the same channel, implementing both $b$ tagging and $c$ tagging (solid line) will have a higher ratio than implementing only $b$ tagging.
To make sure that the ratios are as small as possible and more events survive after the JEP cut, the test cone is chosen as $r=0.3$ for the following analysis.

\begin{figure}[!htb]
\includegraphics[scale=0.37]{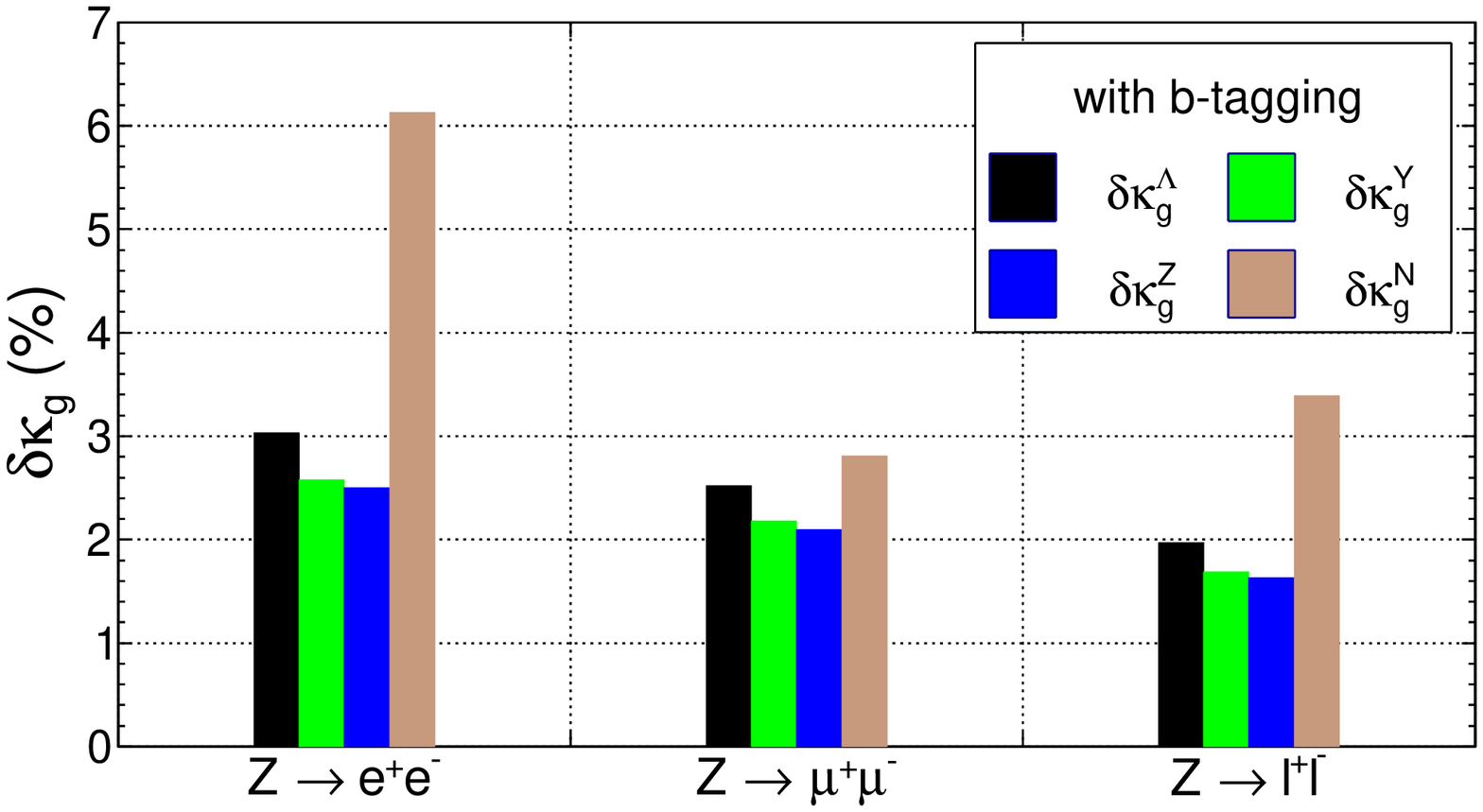}
\includegraphics[scale=0.37]{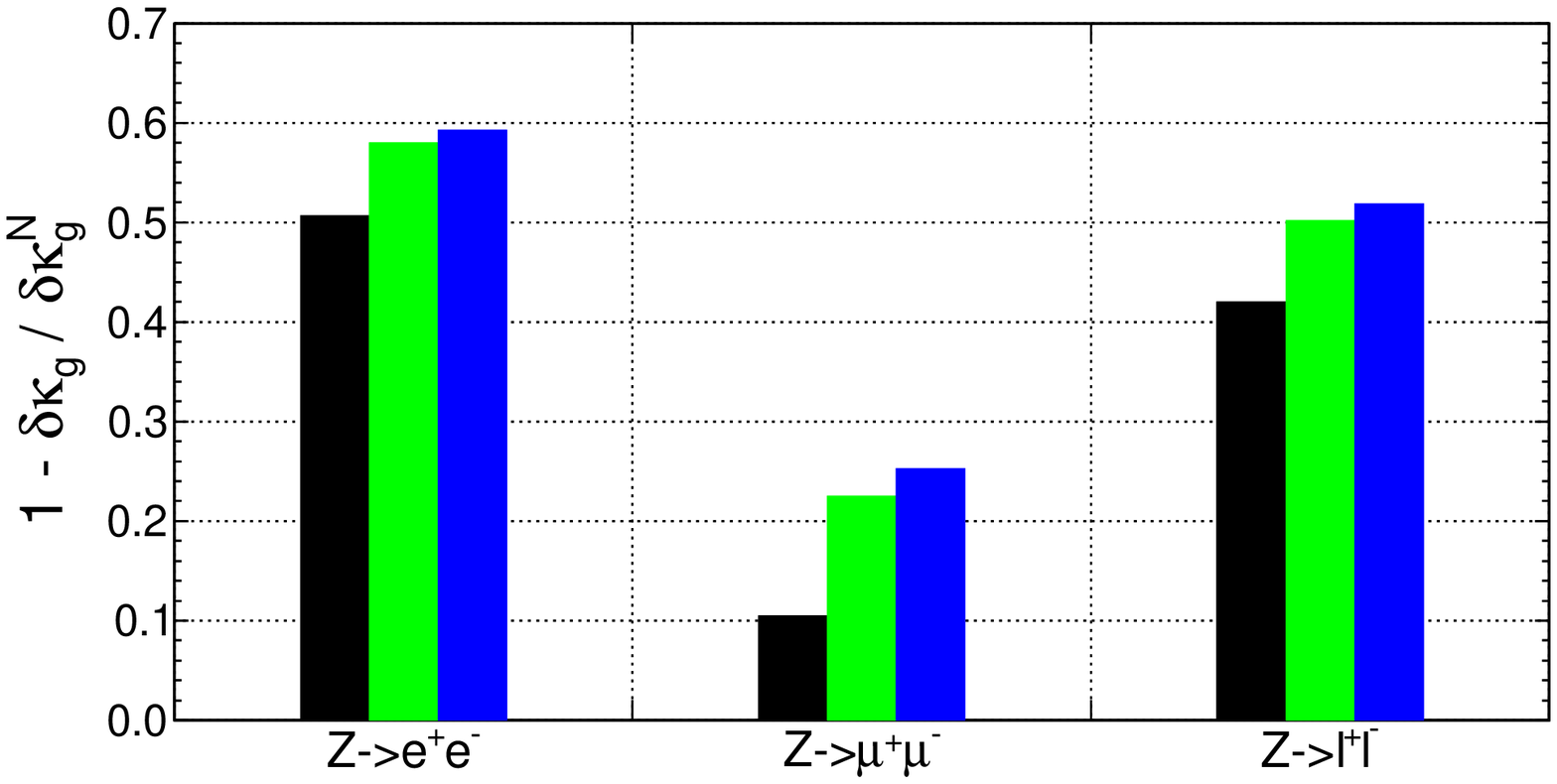}
\caption{\label{FIG:45}
The uncertainties of Higgs boson-gluon effective coupling via different observables (above) and their improvements with respect to $\delta \kappa_g^N$ (below) at test cone $r=0.3$
in different channels of $Z$ boson decay by implementing only $b$ tagging ($\varepsilon_b = 80\%$). }
\end{figure}

\begin{figure}[!htb]
\includegraphics[scale=0.37]{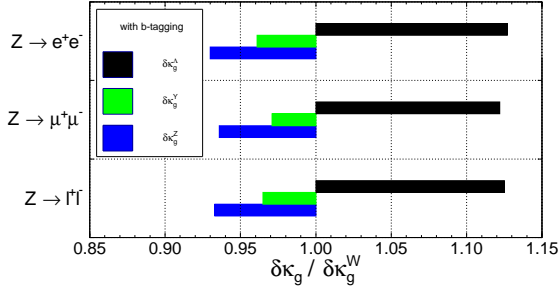}
\caption{\label{FIG:6}
The ratios of Higgs boson-gluon effective coupling uncertainties via different observables with respect to $\delta \kappa_g^W$ at test cone $r=0.3$
in different channels of $Z$ boson decay by implementing only $b$ tagging ($\varepsilon_b = 80\%$). }
\end{figure}

Figure \ref{FIG:45} shows the uncertainties of $\kappa_g$ measurement via different observables (above) and their improvements with respect to $\delta \kappa_g^N$ (below) at test cone $r=0.3$ in different $Z$ boson decay channels by implementing only $b$ tagging.
The uncertainties of $\kappa_g$ via the conventional approach in the electron, muon, and lepton channel are, respectively, 6.1\%, 2.8\%, and 3.4\%.
The uncertainties in the electron channel are much bigger than those in the muon channel, since the bhabha background (final state is $e^+$, $e^-$ and their radiations) has a very large cross section and a sizable number of events still survive after all the kinematic cuts.
However, the JEP cut can effectively remove this kind of background by analyzing the internal structure of jets, since most of these background jets are constituted by only one or a few particles (photon) near the jet axis and given the JEP values very close to both ends.
Therefore, all the new observables will provide remarkable improvements on the conventional approach.
Especially the electron channel, new observables $\Lambda^{N}$, $Y^{N}$, and $Z^{N}$, respectively, get 51\%, 58\%, and 59\% improvements than the conventional approach. When we combine the lepton channels, the uncertainty of $\kappa_g$ can be measured to 1.6\% via the optimized observable $Z^{N}$.

The improvement of the JEP approach comes from the JEP cut and JEP weight. To separate contributions of the two factors, $\delta \kappa_g^W=\sqrt{N^{\rm JEP\_cut}}/2 N_g^{\rm JEP\_cut}$ is used to express the uncertainty of $\kappa_g$ only with a JEP cut. $N_g^{\rm JEP\_cut}$ and $N^{\rm JEP\_cut}$ are, respectively, the number of gluon jets from Higgs boson decay and the total number of jets after the JEP cut.
In Fig.\ref{FIG:6}, it can be seen that observable $\Lambda^{N}$ gets bigger uncertainties of $\kappa_g$ than
$\delta \kappa_g^W$ by about 12\% for the total lepton channel. As we have already analyzed in the last section, due to $\psi_q > \psi_g$,
the ratios $\psi_q / \psi_g$ and $\psi_{\rm BG} / \psi_g$ give heavier weights to quark jets from Higgs boson
decay and background jets. And since $\psi_q$ and $\psi_{\rm BG}$ are bigger than $\psi_g$ by about 13\%,
the observable $\Lambda^{N}$ gets a little bigger uncertainties of $\kappa_g$ than other new observables.
The ratios $(1-\psi_q) / (1-\psi_g)$ and $(1-\psi_{\rm BG}) / (1-\psi_g)$ in observable $Y^{N}$
give lighter weights to quark jets from Higgs boson decay and background jets, so the $\kappa_g$ uncertainties using
observable $Y^{N}$ have about 4\% improvement than $\delta \kappa_g^W$ for the total lepton channel.
After tuning the parameter $a$, the observable $Z^{N}$ is indeed the most optimized one compared to the other observables
and the $\kappa_g$ uncertainties using observable $Z^{N}$ have about 7\% improvement than $\delta \kappa_g^W$
for the total lepton channel. Therefore, this optimized observable $Z^{N}$ is a very promising approach to assist the $\kappa_g$ measurement.

\begin{figure}[!htb]
\includegraphics[scale=0.37]{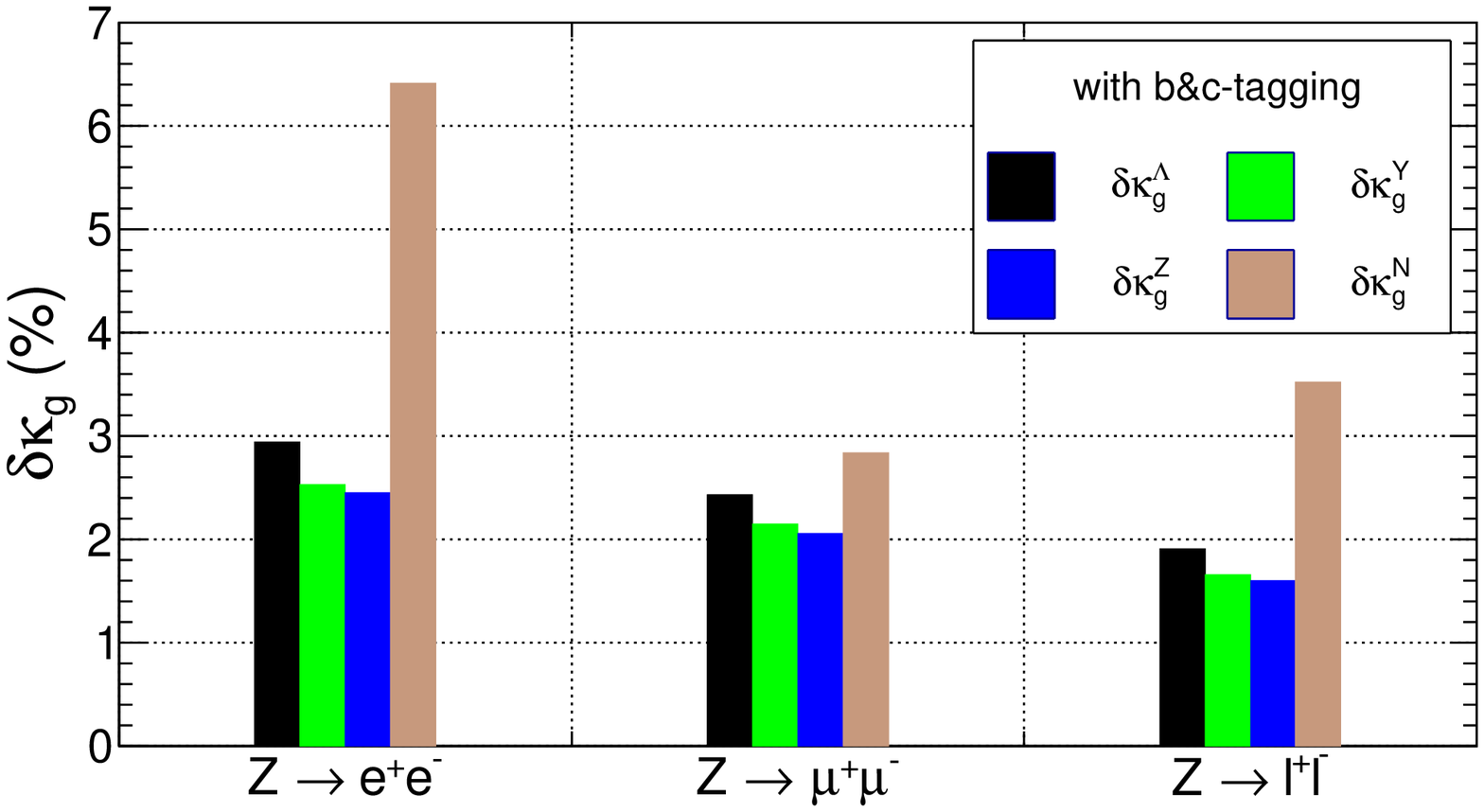}
\includegraphics[scale=0.37]{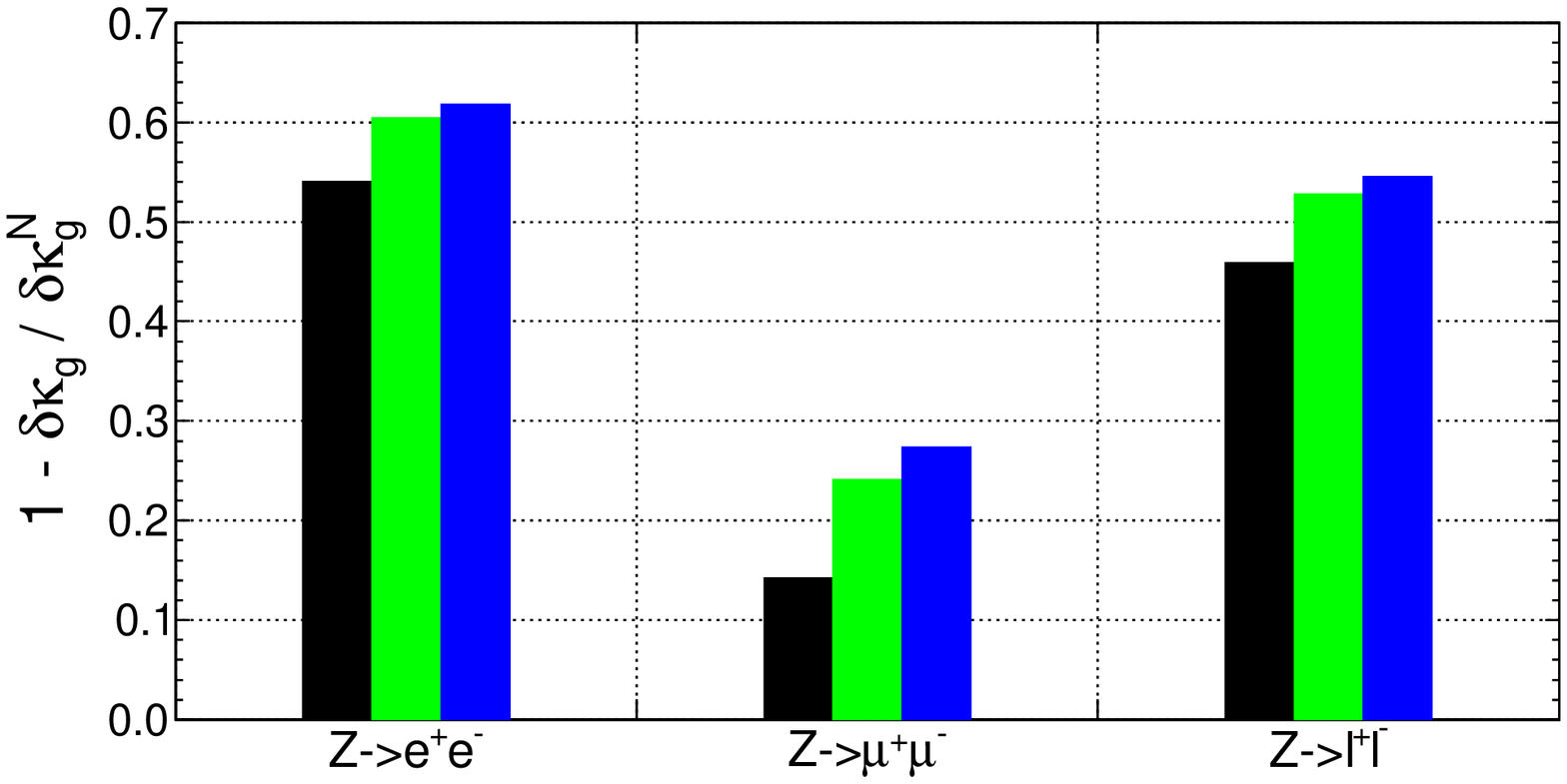}
\caption{\label{FIG:78}
The uncertainties of Higgs boson-gluon effective coupling via different observables (above) and their improvements with respect to $\delta \kappa_g^N$ (below) at test cone $r=0.3$
in different channels of $Z$ boson decay by implementing both $b$ tagging ($\varepsilon_b = 80\%$) and $c$ tagging ($\varepsilon_c = 60\%$). }
\end{figure}

\begin{figure}[!htb]
\includegraphics[scale=0.37]{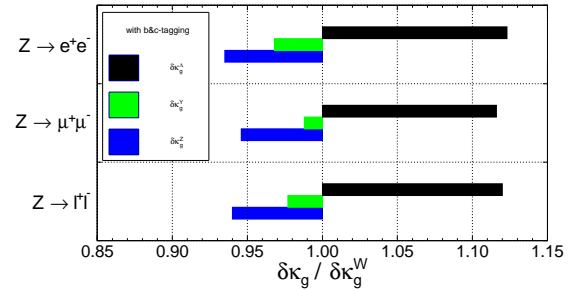}
\caption{\label{FIG:9}
The ratios of Higgs boson-gluon effective coupling uncertainties via different observables with respect to $\delta \kappa_g^W$ at test cone $r=0.3$
in different channels of $Z$ boson decay by implementing both $b$ tagging ($\varepsilon_b = 80\%$) and $c$ tagging ($\varepsilon_c = 60\%$). }
\end{figure}

Figure \ref{FIG:78} presents $\kappa_g$ uncertainties via different observables (above) and their improvements with respect to conventional approach $\delta \kappa_g^N$ (below) including the $b$ and $c$ tagging.
Compared with Fig.\ref{FIG:45}, it can be seen that the uncertainties of $\kappa_g$ via the conventional approach increase by about 5\% after $c$ tagging, although the $c$ tagging can efficiently reduce the contamination of charm jets and the background jets.
This is because the $c$ tagging not only vetoes the charm jets,
but it also excludes some of the gluon jets since its mistag rate for light-quark jet and gluon jet is 7\%.
But the opposite is the uncertainties of $\kappa_g$ via new observables decrease by about 2\% after $c$ tagging.
From the comparison between Fig. \ref{FIG:6} and \ref{FIG:9}, we find that the contributions of JEP weight decrease by about 10\% after $c$ tagging. This means that $c$ tagging can increase contributions of the JEP cut but decrease the contributions of the JEP weight, which leads to the effect from $c$ tagging not being obvious.
In the future, if the mistag rate of $c$ tagging can be improved enough, the $c$ tagging may further help the $\kappa_g$ measurement.

\section{Conclusions}

In this paper, we propose to use the accumulated JEP for the measurement of the Higgs boson-gluon effective coupling.
By using the optimized observable $Z^N$ in the MC simulation at the
future $e^+e^-$ colliders for the center-of-mass energy 250 GeV and integrated luminosity 5 ab$^{-1}$,
the statistical uncertainties of effective coupling $\kappa_g$ can reach about 1.6\% in the channels
of a $Z$ boson decaying to lepton pairs and is totally reduced by about 52\% (45\% from the JEP cut contribution and 7\% from the JEP weight contribution) compared to the relevant $\kappa_g$
uncertainties in the conventional approach. In this work, our MC simulation has not yet included
the template fit, which can further reduce the $\kappa_g$ uncertainties by about 68\%.
If naively implementing this improvement ratio, the $\kappa_g$ uncertainties via the
optimized observable $Z^N$ can be expected to reach 1.1\% after using the template fit.
This will be investigated in detail in our future work.

\vspace{1ex}
This work was supported by the National Natural Science Foundation of China under Grant No. 11675185.
Y.W. is supported by the China Postdoctoral Science Foundation under Grant No. 2016M601134 and an International Postdoctoral Exchange
Fellowship Program between the Office of the National Administrative Committee
of Postdoctoral Researchers of China (ONACPR) and DESY.
X.Z. has received funding from
the European Union's Horizon 2020 research and innovation program as part of the Marie Sk\l{}odowska-Curie
Innovative Training Network MCnetITN3 (Grant Agreement No. 722104).
The authors want to thank Gang Li and Manqi Ruan for helpful discussions and the complete MC simulation events.
\bibliography{jep}

\begin{thebibliography}{31}
\expandafter\ifx\csname natexlab\endcsname\relax\def\natexlab#1{#1}\fi
\expandafter\ifx\csname bibnamefont\endcsname\relax
  \def\bibnamefont#1{#1}\fi
\expandafter\ifx\csname bibfnamefont\endcsname\relax
  \def\bibfnamefont#1{#1}\fi
\expandafter\ifx\csname citenamefont\endcsname\relax
  \def\citenamefont#1{#1}\fi
\expandafter\ifx\csname url\endcsname\relax
  \def\url#1{\texttt{#1}}\fi
\expandafter\ifx\csname urlprefix\endcsname\relax\def\urlprefix{URL }\fi
\providecommand{\bibinfo}[2]{#2}
\providecommand{\eprint}[2][]{\url{#2}}

\bibitem[{\citenamefont{Einhorn}(1993)}]{Einhorn:1993hj}
\bibinfo{author}{\bibfnamefont{M.~B.} \bibnamefont{Einhorn}}, in
  \emph{\bibinfo{booktitle}{{Proceedings of the Conference on Unified Symmetry
  in the Small and in the Large Coral Gables, Florida, January 25-27, 1993}}}
  (\bibinfo{year}{1993}), pp. \bibinfo{pages}{407--420},
  \eprint{hep-ph/9303323}.

\bibitem[{\citenamefont{Kanemura et~al.}(2004)\citenamefont{Kanemura, Okada,
  Senaha, and Yuan}}]{Kanemura:2004mg}
\bibinfo{author}{\bibfnamefont{S.}~\bibnamefont{Kanemura}},
  \bibinfo{author}{\bibfnamefont{Y.}~\bibnamefont{Okada}},
  \bibinfo{author}{\bibfnamefont{E.}~\bibnamefont{Senaha}}, \bibnamefont{and}
  \bibinfo{author}{\bibfnamefont{C.~P.} \bibnamefont{Yuan}},
  \bibinfo{journal}{Phys. Rev.} \textbf{\bibinfo{volume}{D70}},
  \bibinfo{pages}{115002} (\bibinfo{year}{2004}).

\bibitem[{\citenamefont{He et~al.}(2013)\citenamefont{He, Tang, and
  Valencia}}]{He:2013tia}
\bibinfo{author}{\bibfnamefont{X.-G.} \bibnamefont{He}},
  \bibinfo{author}{\bibfnamefont{Y.}~\bibnamefont{Tang}}, \bibnamefont{and}
  \bibinfo{author}{\bibfnamefont{G.}~\bibnamefont{Valencia}},
  \bibinfo{journal}{Phys. Rev.} \textbf{\bibinfo{volume}{D88}},
  \bibinfo{pages}{033005} (\bibinfo{year}{2013}).

\bibitem[{\citenamefont{Moyotl et~al.}(2016)\citenamefont{Moyotl, Chamorro,
  Castilla-Valdez, and Pérez}}]{Moyotl:2016fdk}
\bibinfo{author}{\bibfnamefont{A.}~\bibnamefont{Moyotl}},
  \bibinfo{author}{\bibfnamefont{S.}~\bibnamefont{Chamorro}},
  \bibinfo{author}{\bibfnamefont{H.}~\bibnamefont{Castilla-Valdez}},
  \bibnamefont{and} \bibinfo{author}{\bibfnamefont{M.~A.} \bibnamefont{Pérez}}
  (\bibinfo{year}{2016}), \eprint{1610.06299}.

\bibitem[{\citenamefont{Baek and Yuan}(2017)}]{Baek:2017kxh}
\bibinfo{author}{\bibfnamefont{S.}~\bibnamefont{Baek}} \bibnamefont{and}
  \bibinfo{author}{\bibfnamefont{X.-B.} \bibnamefont{Yuan}},
  \bibinfo{journal}{Phys. Lett.} \textbf{\bibinfo{volume}{B774}},
  \bibinfo{pages}{662} (\bibinfo{year}{2017}).

\bibitem[{\citenamefont{Hou and Kikuchi}(2017)}]{Hou:2017vvp}
\bibinfo{author}{\bibfnamefont{W.-S.} \bibnamefont{Hou}} \bibnamefont{and}
  \bibinfo{author}{\bibfnamefont{M.}~\bibnamefont{Kikuchi}},
  \bibinfo{journal}{Phys. Rev.} \textbf{\bibinfo{volume}{D96}},
  \bibinfo{pages}{015033} (\bibinfo{year}{2017}).

\bibitem[{\citenamefont{Kanemura et~al.}(2017)\citenamefont{Kanemura, Kikuchi,
  Sakurai, and Yagyu}}]{Kanemura:2017wtm}
\bibinfo{author}{\bibfnamefont{S.}~\bibnamefont{Kanemura}},
  \bibinfo{author}{\bibfnamefont{M.}~\bibnamefont{Kikuchi}},
  \bibinfo{author}{\bibfnamefont{K.}~\bibnamefont{Sakurai}}, \bibnamefont{and}
  \bibinfo{author}{\bibfnamefont{K.}~\bibnamefont{Yagyu}},
  \bibinfo{journal}{Phys. Rev.} \textbf{\bibinfo{volume}{D96}},
  \bibinfo{pages}{035014} (\bibinfo{year}{2017}).

\bibitem[{\citenamefont{Paßehr and Weiglein}(2017)}]{Passehr:2017ufr}
\bibinfo{author}{\bibfnamefont{S.}~\bibnamefont{Paßehr}} \bibnamefont{and}
  \bibinfo{author}{\bibfnamefont{G.}~\bibnamefont{Weiglein}}
  (\bibinfo{year}{2017}), \eprint{1705.07909}.

\bibitem[{\citenamefont{Aad et~al.}(2016{\natexlab{a}})}]{Khachatryan:2016vau}
\bibinfo{author}{\bibfnamefont{G.}~\bibnamefont{Aad}} \bibnamefont{et~al.}
  (\bibinfo{collaboration}{ATLAS, CMS}), \bibinfo{journal}{JHEP}
  \textbf{\bibinfo{volume}{08}}, \bibinfo{pages}{045}
  (\bibinfo{year}{2016}{\natexlab{a}}).

\bibitem[{\citenamefont{Collaboration}(2018)}]{CMS:2018lkl}
\bibinfo{author}{\bibfnamefont{C.}~\bibnamefont{Collaboration}}
  (\bibinfo{collaboration}{CMS}), \emph{\bibinfo{title}{{Combined measurements
  of the Higgs boson's couplings at $\sqrt{s}=13$ TeV}}}
  (\bibinfo{year}{2018}).

\bibitem[{\citenamefont{Peskin}(2012)}]{Peskin:2012we}
\bibinfo{author}{\bibfnamefont{M.~E.} \bibnamefont{Peskin}}
  (\bibinfo{year}{2012}), \eprint{1207.2516}.

\bibitem[{\citenamefont{Peskin}(2013)}]{Peskin:2013xra}
\bibinfo{author}{\bibfnamefont{M.~E.} \bibnamefont{Peskin}}, in
  \emph{\bibinfo{booktitle}{{Proceedings, 2013 Community Summer Study on the
  Future of U.S. Particle Physics: Snowmass on the Mississippi (CSS2013):
  Minneapolis, MN, USA, July 29-August 6, 2013}}} (\bibinfo{year}{2013}),
  \eprint{1312.4974}.

\bibitem[{CEP(2015{\natexlab{a}})}]{CEPC-SPPCStudyGroup:2015csa}
\emph{\bibinfo{title}{{CEPC-SPPC Preliminary Conceptual Design Report. 1.
  Physics and Detector}}} (\bibinfo{year}{2015}{\natexlab{a}}),
  \eprint{IHEP-CEPC-DR-2015-01, IHEP-TH-2015-01, IHEP-EP-2015-01}.

\bibitem[{CEP(2015{\natexlab{b}})}]{CEPC-SPPCStudyGroup:2015esa}
\emph{\bibinfo{title}{{CEPC-SPPC Preliminary Conceptual Design Report. 2.
  Accelerator}}} (\bibinfo{year}{2015}{\natexlab{b}}),
  \eprint{IHEP-CEPC-DR-2015-01, IHEP-AC-2015-01}.

\bibitem[{\citenamefont{Mo et~al.}(2016)\citenamefont{Mo, Li, Ruan, and
  Lou}}]{Mo:2015mza}
\bibinfo{author}{\bibfnamefont{X.}~\bibnamefont{Mo}},
  \bibinfo{author}{\bibfnamefont{G.}~\bibnamefont{Li}},
  \bibinfo{author}{\bibfnamefont{M.-Q.} \bibnamefont{Ruan}}, \bibnamefont{and}
  \bibinfo{author}{\bibfnamefont{X.-C.} \bibnamefont{Lou}},
  \bibinfo{journal}{Chin. Phys.} \textbf{\bibinfo{volume}{C40}},
  \bibinfo{pages}{033001} (\bibinfo{year}{2016}).

\bibitem[{\citenamefont{Bicer et~al.}(2014)}]{Gomez-Ceballos:2013zzn}
\bibinfo{author}{\bibfnamefont{M.}~\bibnamefont{Bicer}} \bibnamefont{et~al.}
  (\bibinfo{collaboration}{TLEP Design Study Working Group}),
  \bibinfo{journal}{JHEP} \textbf{\bibinfo{volume}{01}}, \bibinfo{pages}{164}
  (\bibinfo{year}{2014}).

\bibitem[{\citenamefont{Barletta et~al.}(2014)\citenamefont{Barletta,
  Battaglia, Klute, Mangano, Prestemon, Rossi, and Skands}}]{Barletta:2014vea}
\bibinfo{author}{\bibfnamefont{W.}~\bibnamefont{Barletta}},
  \bibinfo{author}{\bibfnamefont{M.}~\bibnamefont{Battaglia}},
  \bibinfo{author}{\bibfnamefont{M.}~\bibnamefont{Klute}},
  \bibinfo{author}{\bibfnamefont{M.}~\bibnamefont{Mangano}},
  \bibinfo{author}{\bibfnamefont{S.}~\bibnamefont{Prestemon}},
  \bibinfo{author}{\bibfnamefont{L.}~\bibnamefont{Rossi}}, \bibnamefont{and}
  \bibinfo{author}{\bibfnamefont{P.}~\bibnamefont{Skands}},
  \bibinfo{journal}{Nucl. Instrum. Meth.} \textbf{\bibinfo{volume}{A764}},
  \bibinfo{pages}{352} (\bibinfo{year}{2014}).

\bibitem[{\citenamefont{Benedikt and Zimmermann}(2016)}]{Benedikt:2016vzy}
\bibinfo{author}{\bibfnamefont{M.}~\bibnamefont{Benedikt}} \bibnamefont{and}
  \bibinfo{author}{\bibfnamefont{F.}~\bibnamefont{Zimmermann}},
  \bibinfo{journal}{PoS} \textbf{\bibinfo{volume}{LeptonPhoton2015}},
  \bibinfo{pages}{052} (\bibinfo{year}{2016}).

\bibitem[{\citenamefont{Behnke et~al.}(2013)\citenamefont{Behnke, Brau, Foster,
  Fuster, Harrison, Paterson, Peskin, Stanitzki, Walker, and
  Yamamoto}}]{Behnke:2013xla}
\bibinfo{author}{\bibfnamefont{T.}~\bibnamefont{Behnke}},
  \bibinfo{author}{\bibfnamefont{J.~E.} \bibnamefont{Brau}},
  \bibinfo{author}{\bibfnamefont{B.}~\bibnamefont{Foster}},
  \bibinfo{author}{\bibfnamefont{J.}~\bibnamefont{Fuster}},
  \bibinfo{author}{\bibfnamefont{M.}~\bibnamefont{Harrison}},
  \bibinfo{author}{\bibfnamefont{J.~M.} \bibnamefont{Paterson}},
  \bibinfo{author}{\bibfnamefont{M.}~\bibnamefont{Peskin}},
  \bibinfo{author}{\bibfnamefont{M.}~\bibnamefont{Stanitzki}},
  \bibinfo{author}{\bibfnamefont{N.}~\bibnamefont{Walker}}, \bibnamefont{and}
  \bibinfo{author}{\bibfnamefont{H.}~\bibnamefont{Yamamoto}}
  (\bibinfo{year}{2013}), \eprint{1306.6327}.

\bibitem[{\citenamefont{Adolphsen et~al.}(2013)\citenamefont{Adolphsen, Barone,
  Barish, Buesser, Burrows, Carwardine, Clark, Mainaud~Durand, Dugan, Elsen
  et~al.}}]{Adolphsen:2013kya}
\bibinfo{author}{\bibfnamefont{C.}~\bibnamefont{Adolphsen}},
  \bibinfo{author}{\bibfnamefont{M.}~\bibnamefont{Barone}},
  \bibinfo{author}{\bibfnamefont{B.}~\bibnamefont{Barish}},
  \bibinfo{author}{\bibfnamefont{K.}~\bibnamefont{Buesser}},
  \bibinfo{author}{\bibfnamefont{P.}~\bibnamefont{Burrows}},
  \bibinfo{author}{\bibfnamefont{J.}~\bibnamefont{Carwardine}},
  \bibinfo{author}{\bibfnamefont{J.}~\bibnamefont{Clark}},
  \bibinfo{author}{\bibfnamefont{H.}~\bibnamefont{Mainaud~Durand}},
  \bibinfo{author}{\bibfnamefont{G.}~\bibnamefont{Dugan}},
  \bibinfo{author}{\bibfnamefont{E.}~\bibnamefont{Elsen}}, \bibnamefont{et~al.}
  (\bibinfo{year}{2013}), \eprint{1306.6328}.

\bibitem[{\citenamefont{Abramowicz et~al.}(2013)}]{Behnke:2013lya}
\bibinfo{author}{\bibfnamefont{H.}~\bibnamefont{Abramowicz}}
  \bibnamefont{et~al.} (\bibinfo{year}{2013}), \eprint{1306.6329}.

\bibitem[{\citenamefont{Bai}(2017)}]{YuBai}
\bibinfo{author}{\bibfnamefont{Y.}~\bibnamefont{Bai}}
  (\bibinfo{collaboration}{CEPC Working Group}),
  \emph{\bibinfo{title}{{Measurements of the decay branching fraction of $H\to
  b\bar b/c\bar c/gg$ at CEPC (CEPC Note) (to be published)}}}
  (\bibinfo{year}{2017}).

\bibitem[{\citenamefont{Aad et~al.}(2016{\natexlab{b}})}]{Aad:2015gba}
\bibinfo{author}{\bibfnamefont{G.}~\bibnamefont{Aad}} \bibnamefont{et~al.}
  (\bibinfo{collaboration}{ATLAS}), \bibinfo{journal}{Eur. Phys. J.}
  \textbf{\bibinfo{volume}{C76}}, \bibinfo{pages}{6}
  (\bibinfo{year}{2016}{\natexlab{b}}).

\bibitem[{\citenamefont{Rentala et~al.}(2013)\citenamefont{Rentala, Vignaroli,
  Li, Li, and Yuan}}]{Rentala:2013uaa}
\bibinfo{author}{\bibfnamefont{V.}~\bibnamefont{Rentala}},
  \bibinfo{author}{\bibfnamefont{N.}~\bibnamefont{Vignaroli}},
  \bibinfo{author}{\bibfnamefont{H.-n.} \bibnamefont{Li}},
  \bibinfo{author}{\bibfnamefont{Z.}~\bibnamefont{Li}}, \bibnamefont{and}
  \bibinfo{author}{\bibfnamefont{C.~P.} \bibnamefont{Yuan}},
  \bibinfo{journal}{Phys. Rev.} \textbf{\bibinfo{volume}{D88}},
  \bibinfo{pages}{073007} (\bibinfo{year}{2013}).

\bibitem[{\citenamefont{Agrawal and Rentala}(2014)}]{Agrawal:2013hya}
\bibinfo{author}{\bibfnamefont{P.}~\bibnamefont{Agrawal}} \bibnamefont{and}
  \bibinfo{author}{\bibfnamefont{V.}~\bibnamefont{Rentala}},
  \bibinfo{journal}{JHEP} \textbf{\bibinfo{volume}{05}}, \bibinfo{pages}{098}
  (\bibinfo{year}{2014}).

\bibitem[{\citenamefont{Chivukula et~al.}(2015)\citenamefont{Chivukula,
  Simmons, and Vignaroli}}]{Chivukula:2015tca}
\bibinfo{author}{\bibfnamefont{R.~S.} \bibnamefont{Chivukula}},
  \bibinfo{author}{\bibfnamefont{E.~H.} \bibnamefont{Simmons}},
  \bibnamefont{and}
  \bibinfo{author}{\bibfnamefont{N.}~\bibnamefont{Vignaroli}}, in
  \emph{\bibinfo{booktitle}{{Sakata Memorial KMI Workshop on Origin of Mass and
  Strong Coupling Gauge Theories (SCGT15) Nagoya, Japan, March 3-6, 2015}}}
  (\bibinfo{year}{2015}), \eprint{1507.06522}.

\bibitem[{\citenamefont{Li et~al.}(2011)\citenamefont{Li, Li, and
  Yuan}}]{Li:2011hy}
\bibinfo{author}{\bibfnamefont{H.-n.} \bibnamefont{Li}},
  \bibinfo{author}{\bibfnamefont{Z.}~\bibnamefont{Li}}, \bibnamefont{and}
  \bibinfo{author}{\bibfnamefont{C.~P.} \bibnamefont{Yuan}},
  \bibinfo{journal}{Phys. Rev. Lett.} \textbf{\bibinfo{volume}{107}},
  \bibinfo{pages}{152001} (\bibinfo{year}{2011}).

\bibitem[{\citenamefont{Mo and Li}(2017)}]{XinMo}
\bibinfo{author}{\bibfnamefont{X.}~\bibnamefont{Mo}} \bibnamefont{and}
  \bibinfo{author}{\bibfnamefont{G.}~\bibnamefont{Li}}
  (\bibinfo{collaboration}{CEPC Working Group}),
  \emph{\bibinfo{title}{{Generated Sample Stauts for CEPC Simulation Studies
  (CEPC Note) (to be published)}}} (\bibinfo{year}{2017}).

\bibitem[{\citenamefont{Yan et~al.}(2016)\citenamefont{Yan, Watanuki, Fujii,
  Ishikawa, Jeans, Strube, Tian, and Yamamoto}}]{Yan:2016xyx}
\bibinfo{author}{\bibfnamefont{J.}~\bibnamefont{Yan}},
  \bibinfo{author}{\bibfnamefont{S.}~\bibnamefont{Watanuki}},
  \bibinfo{author}{\bibfnamefont{K.}~\bibnamefont{Fujii}},
  \bibinfo{author}{\bibfnamefont{A.}~\bibnamefont{Ishikawa}},
  \bibinfo{author}{\bibfnamefont{D.}~\bibnamefont{Jeans}},
  \bibinfo{author}{\bibfnamefont{J.}~\bibnamefont{Strube}},
  \bibinfo{author}{\bibfnamefont{J.}~\bibnamefont{Tian}}, \bibnamefont{and}
  \bibinfo{author}{\bibfnamefont{H.}~\bibnamefont{Yamamoto}},
  \bibinfo{journal}{Phys. Rev.} \textbf{\bibinfo{volume}{D94}},
  \bibinfo{pages}{113002} (\bibinfo{year}{2016}).

\bibitem[{\citenamefont{Cacciari et~al.}(2012)\citenamefont{Cacciari, Salam,
  and Soyez}}]{Cacciari:2011ma}
\bibinfo{author}{\bibfnamefont{M.}~\bibnamefont{Cacciari}},
  \bibinfo{author}{\bibfnamefont{G.~P.} \bibnamefont{Salam}}, \bibnamefont{and}
  \bibinfo{author}{\bibfnamefont{G.}~\bibnamefont{Soyez}},
  \bibinfo{journal}{Eur. Phys. J.} \textbf{\bibinfo{volume}{C72}},
  \bibinfo{pages}{1896} (\bibinfo{year}{2012}).

\bibitem[{\citenamefont{Ruan}(2017)}]{ManqiRUAN}
\bibinfo{author}{\bibfnamefont{M.}~\bibnamefont{Ruan}}
  (\bibinfo{collaboration}{CEPC-SPPC Study group}),
  \emph{\bibinfo{title}{{Status and Updates from CEPC Simulation-Detector
  optimization}}},
  \bibinfo{howpublished}{\url{http://ias.ust.hk/program/shared_doc/2017/
  201701hep/HEP_20170124_Manqi_Ruan.pdf}} (\bibinfo{year}{2017}).

\end{thebibliography}

\end{document}